# Reliably determining which genes have a high posterior probability of differential expression

## A microarray application of decision-theoretic multiple testing

**David R. Bickel**

August 14, 2003; modified February 27, 2004

*Office of Biostatistics and Bioinformatics*
*Medical College of Georgia*
*Augusta, GA 30912-4900*
*dbickel@mcg.edu*
*www.davidbickel.com*

## Abstract

Microarray data are often used to determine which genes are differentially expressed between groups, for example, between treatment and control groups. There are methods of determining which genes have a high probability of differential expression, but those methods depend on the estimation of probability densities. Theoretical results have shown such estimation to be unreliable when high-probability genes are identified.

The genes that are probably differentially expressed can be found using decision theory instead of density estimation. Simulations show that the proposed decision-theoretic method is much more reliable than a density-estimation method. The proposed method is used to determine which genes to consider differentially expressed between patients with different types of cancer.

The proposed method determines which genes have a high probability of differential expression. It can be applied to data sets that have replicate microarrays in each of two or more groups of patients or experiments.

**Introduction**

The widespread use of microarray technology has been generating much interest in methods for testing multiple hypotheses. A typical microarray experiment has measurements of the expression levels of thousands of genes. In many cases, the investigator wishes to make inferences about each of the genes, for example, whether or not each gene is expressed differently in a treatment group and in a control group. This problem has been approached statistically by testing a null hypothesis for each gene, that the gene is not differentially expressed between the two groups of interest. The rejection of the null hypothesis associated with a gene then means that the gene is considered to be differentially expressed; this is called a *discovery* of differential expression. A discovery is true when the gene expression is different in the population, and is false when the observed difference is due to chance variation. In statistical terms, a true discovery is the rejection of a false null hypothesis and a false discovery is the rejection of a true null hypothesis. The practice of rejecting all null hypotheses with a p-value less than a significance level $\alpha$, such as 0.05, is very misleading for microarray data since a portion $\alpha$ of all genes would be then considered differentially expressed when there is no difference between the populations of the two samples. Hence, special statistical methods for the testing of multiple hypotheses are called for. In one of the earliest applications of such methods to microarray data, Dudoit *et al*. (2002) took the conservative approach of ensuring that the probability of making any false discovery was strictly limited. However, many investigators now realize that, for their purposes, it is unnecessary and often impractical to attempt to avoid making even a single false discovery since allowing a few false discoveries can lead to many more true discoveries. Thus, many less conservative multiple hypothesis methods have been applied to the analysis of microarray data, using different criteria for deciding which null hypotheses to reject. Three closely related criteria that have been applied to microarray analysis are:

1. *Control of a false discovery rate*. As many genes as possible are considered differentially expressed while holding a false discovery rate below a fixed threshold. Informally, the false discovery rate is the proportion of discoveries of gene expression that are false, i.e., the ratio of the number of genes considered differentially expressed that are not really differentially expressed to the total number of genes considered differentially expressed. For example, if 200 genes are considered differentially expressed and the FDR is 10%, then only 180 of those genes are really differentially expressed. Different mathematical definitions of the false discovery rate include the FDR (Benjamini and Hochberg 1995), the positive FDR (pFDR; Storey 2002a), and the decisive FDR (dFDR; Bickel 2003, 2004), denoted by $\varepsilon$, $Q$, and $\Delta$, respectively.

2. *Decision-theoretic optimization*. Genes are considered differentially expressed such that the net benefit minus the net cost is maximized, based on known (Storey 2002b) or estimated (Müller *et al*. 2004; Bickel 2003) probability distributions.

3. *Achievement of a posterior probability*. A gene is considered differentially expressed if it has a sufficiently high posterior probability of differential expression (Efron et al. 2001; Genovese and Wasserman 2002; Müller *et al*. 2004). Genovese and Wasserman (2002) point out a potentially serious disadvantage of these methods in the nonparametric case: they require the estimation of a ratio of probability densities, estimation that is unreliable in the distribution tail regions.

It will be seen that the goals of the decision-theoretic criterion and the posterior probability criterion are identical under certain conditions. This identity has the practical implication that the posterior probability criterion can be implemented without estimating a ratio of probability densities. Potential areas of application include not only the detection of differentially expressed genes, but also other situations in which a large number of hypotheses are tested, for example, the identification of exonic splicing enhancers, the genetic dissection of transcriptional regulation, and the finding of binding sites of transcriptional regulators (Storey and Tibshirani 2003). Methods of multiple hypothesis testing will probably also aid in the analysis of proteomic data, as in experiments designed to detect differences in protein levels between different groups of subjects since such experiments measure the abundances of many proteins simultaneously.

## Theory

### Formulation in terms of false discoveries and false nondiscoveries

A discovery is made whenever a null hypothesis is rejected, which occurs when a test statistic falls within the rejection region, $\Gamma$, a subset of the test statistic space. A *true discovery* is a rejection of a false null hypothesis and a *false discovery* is the rejection of a true null hypothesis. Let $c$ be the cost of each false discovery and $c'$ the cost of each false nondiscovery; $c > 0$, $c' > 0$. Then the optimization parameter, $p$, is defined by $p = 1/(c/c' + 1)$. Storey (2002b) studied *Bayes' error*,

$$\text{BE}(\Gamma) \equiv (1 - p) \Pr(t_i \in \Gamma, H_i = 0) + p \Pr(t_i \notin \Gamma, H_i = 1), \tag{1}$$

where $t_i$ is the statistic associated with the $i$th hypothesis test and where $H_i = 0$ or $H_i = 1$ if the $i$th null hypothesis is true or false, respectively. The decision-theoretic goal is to find the rejection region that minimizes $\text{BE}(\Gamma)$. For ease of notation, consider the rejection region $\Gamma = \{t : t \geq \tau\}$. An example is rejecting all null hypothesis with p-values below some significance level, in which case $t, \tau \in [-1, 0]$. If the distribution of test statistics is $F_0(t)$ under the null hypothesis and $F_1(\tau)$ under the alternative hypothesis, then

$$\Pr(t_i \notin \Gamma \mid H_i) = \Pr(t_i < \tau \mid H_i) = (1 - H_i) F_0(\tau) + H_i F_1(\tau) \tag{2}$$

and, from Bayes' Rule,

$$\mathrm{BE}(\Gamma) \equiv B(\tau) = (1-p)\Pr(t_i \geq \tau, H_i = 0) + p\Pr(t_i < \tau, H_i = 1) = \\ (1-p)\Pr(H_i = 0)(1 - F_0(\tau)) + p\Pr(H_i = 1)F_1(\tau). \tag{3}$$

It is assumed that the test statistics are drawn from the mixed distribution $F(\tau) = \pi_0 F_0(\tau) + \pi_1 F_1(\tau)$, with $\pi_0 \equiv \Pr(H_i = 0)$ and $\pi_1 \equiv \Pr(H_i = 1)$ and with densities $f$, $f_0$, and $f_1$ corresponding to $F$, $F_0$, and $F_1$, respectively. Denoting by $\tilde{\tau}$ the value of $\tau$ at which $B(\tau)$ is maximized, the differentiation of $B(\tau)$ yields

$$p = \frac{\pi_0 f_0(\tilde{\tau})}{\pi_0 f_0(\tilde{\tau}) + \pi_1 f_1(\tilde{\tau})} = \frac{\pi_0 f_0(\tilde{\tau})}{f(\tilde{\tau})}, \tag{4}$$

in agreement with Storey (2002b), since each $H_i$ is independent of $\tau$. Again applying Bayes' Rule,

$$p = \frac{\Pr(H_i = 0)\Pr(t_i \in [\tilde{\tau}, \tilde{\tau} + d\tau) \mid H_i = 0)}{\Pr(t_i \in [\tilde{\tau}, \tilde{\tau} + d\tau))} = \\ \Pr(H_i = 0 \mid t_i \in [\tilde{\tau}, \tilde{\tau} + d\tau)) \equiv \Pr(H = 0 \mid t = \tilde{\tau}), \tag{5}$$

i.e., $p$ is the posterior probability that $H_i = 0$ if the test statistic is on the border of the optimal rejection region. Given the reasonable assumption that $\Pr(H = 0 \mid t = t')$ is a monotonic function of $t'$,

$$\forall_{t' \geq \tilde{\tau}} \Pr(H = 0 \mid t = t') \leq p. \tag{6}$$

It follows that the Bayesian practice of rejecting null hypotheses with values of $\Pr(H = 0 \mid t = t')$ less than or equal to some threshold $p$ is equivalent to rejecting null hypotheses according to the decision-theoretic approach with optimization parameter $p$. Müller *et al.* (2004) also noticed this connection between decision-theoretic optimization and using a threshold of a posterior probability.

**Formulation in terms of true and false discoveries**

The decision-theoretic optimization has equivalently been formulated without reference to nondiscoveries by considering the cost, $c$, of each false discovery and the benefit, $b$, of each true discovery (Bickel 2003). The expected net benefit minus the expected net cost is maximized when the *Bayes' desirability*, defined by

$$\text{BD}(\Gamma) \equiv p \Pr(t_i \in \Gamma, H_i = 1) - (1-p)\Pr(t_i \in \Gamma, H_i = 0), \tag{7}$$

is maximized. Here, the optimization parameter is $p = 1/(c/b + 1)$. Since, according to Bayes' Rule, $\text{BD}(\Gamma) = -\text{BE}(\Gamma) + p\Pr(H_i = 1)$, the rejection region that minimizes $\text{BE}(\Gamma)$ also maximizes $\text{BD}(\Gamma)$, i.e., $\operatorname{argmin}_\Gamma \text{BE}(\Gamma) = \operatorname{argmax}_\Gamma \text{BD}(\Gamma)$.

**False discovery rates**

This optimization parameter, $p$, can be related to false discovery rates using a result of Storey (2002b):

$$\Pr(H = 0 \mid t \in \Gamma) = \Delta \approx Q, \tag{8}$$

where the approximation holds for weak dependence between test statistics and for a large number of hypotheses. Again assuming monotonicity, Eq. (8) implies that the dFDR, the expected number of false discoveries divided by the expected number of all discoveries, is

$$\Delta = \Pr(H = 0 \mid t > \tau) < \Pr(H = 0 \mid t = \tau), \tag{9}$$

which Genovese and Wasserman (2002) provide as an approximate result for the FDR since $\varepsilon \approx \Delta$ under certain conditions. For the optimal rejection region, we then have

$$\tilde{\Delta} = \Pr(H = 0 \mid t > \tilde{\tau}) < p, \tag{10}$$

from Eq. (5), where $\tilde{\Delta}$ is the dFDR when the net loss is minimal. This inequality was originally reported in terms of the equivalent true/false discovery formulation: $\tilde{\Delta} < 1/(1 + c/b)$ (Bickel 2003). In fact, selecting $c/b$ such that $p$ was equal to a given upper bound for $\tilde{\Delta}$ (Bickel 2003) implicitly rejected null hypotheses that had posterior probabilities less than $p$.

**Determination of which null hypotheses are improbable**

The optimal rejection region can be estimated by maximizing estimates of the Bayes' desirability. Without loss of generality, we again consider rejection regions of the form $\Gamma = \{t : t \geq \tau\}$. Then Eq. (7) can be written as

$$\text{BD}(\Gamma) \equiv \delta(\tau) = p \Pr(t_i \geq \tau, H_i = 1) - (1-p) \Pr(t_i \geq \tau, H_i = 0), \tag{11}$$

or, using Bayes' rule,

$$\delta(\tau) = p \Pr(t_i \geq \tau) - \pi_0 \Pr(t_i \geq \tau \mid H_i = 0), \tag{12}$$

which is naturally estimated by

$$\hat{\delta}(\tau) = p\left(1 - \hat{F}(\tau)\right) - \hat{\pi}_0\left(1 - \hat{F}_0(\tau)\right), \tag{13}$$

where $\hat{\pi}_0$, $\hat{F}_0(\tau)$, and $\hat{F}(\tau)$ are estimators of $\pi_0$, $F_0(\tau)$, and $F(\tau)$, respectively. Alternately, the Bayes' error (3),

$$B(\tau) = p \Pr(t_i < \tau) - \pi_0 \Pr(t_i < \tau \mid H_i = 0) + (1-p)\pi_0, \tag{14}$$

could be similarly estimated by

$$\hat{B}(\tau) = p\,\hat{F}(\tau) + \hat{\pi}_0\left(1 - \hat{F}_0(\tau)\right) - p\,\hat{\pi}_0, \tag{15}$$

but the extra term is unnecessary for optimization purposes, so Eq. (13) is more convenient since it yields the same optimal rejection region. When the null distribution is known, $\hat{F}_0(\tau)$ should be replaced with $F_0(\tau)$; for example, if $t$ is the additive inverse of a p-value, then $F_0(\tau) = 1 - |\tau|$, as per Storey (2002a) and Genovese and Wasserman (2002). Estimators of false discovery rates used by Efron et al. (2001), Genovese and Wasserman (2002), and Storey (2002a) are also based on $\hat{\pi}_0$, $F_0(\tau)$ or $\hat{F}_0(\tau)$, and $\hat{F}(\tau)$, taking the form

$$\hat{\Delta}(\tau) \equiv \frac{\hat{\pi}_0\left(1 - \hat{F}_0(\tau)\right)}{1 - \hat{F}(\tau)}, \tag{16}$$

again replacing $\hat{F}_0(\tau)$ by $F_0(\tau)$, if known. ($\hat{\Delta}(\tau)$ estimates the FDR or pFDR under the independence or weak dependence of test statistics (Storey 2002b) and estimates the dFDR under more general conditions (Bickel 2003, 2004).) Thus, the algorithms of Efron et al. (2001), Genovese and Wasserman (2002), and Storey (2002a) that were used to compute $\hat{\pi}_0$, $\hat{F}_0(\tau)$, and $\hat{F}(\tau)$ can be applied to the estimation of $\hat{\delta}(\tau)$ as well as $\hat{\Delta}(\tau)$; the estimates are related by $\hat{\delta}(\tau) = (p - \hat{\Delta}(\tau))(1 - \hat{F}(\tau))$. Then the optimal rejection region is estimated by $\{t : t \geq \hat{\tau}\}$, where $\hat{\tau} \equiv \arg\max_{\tau \in \{T_1, T_2, \ldots, T_m\}} \hat{\delta}(\tau)$, with $\hat{\delta}(\tau)$ computed from Eq. (13) for each of the observed test statistics, $\{T_1, T_2, \ldots, T_m\}$. Since null hypotheses of test statistics that fall in the rejection region have a posterior probability that is less than or equal to $p$, as seen above (6), the estimated rejection region determines which null hypotheses to consider improbable enough to reject. Thus, unlike previous methods of rejecting null hypotheses based on their posterior probabilities (Efron et al. 2001; Genovese and Wasserman 2002), the approach taken herein does not require the estimation of densities or of their ratios.

One way to compute $\hat{\pi}_0$, $\hat{F}_0(\tau)$, and $\hat{F}(\tau)$ for Eq. (13) is that described by Bickel (2003), a variant of the algorithms of Efron et al. (2001) and Storey (2002a). Using the absolute value of the two-sample, unequal variance t-statistic as $t$, $\hat{F}(\tau)$ was computed as the proportion of test statistics less than or equal to $\tau$. The columns of the variable-by-case (e.g., gene-by-microarray) matrix were randomly permuted a large number of times, generating null test statistics for each of $m$ variables. Then $\hat{F}_0(\tau)$ was the proportion of null test statistics less than or equal to $\tau$. Using the same methods of estimating cumulative distributions, $\pi_0$ was estimated

$$\hat{\pi}_0(\lambda) = \hat{F}(\lambda) / \hat{F}_0(\lambda), \tag{17}$$

where $\lambda$ was chosen such that approximately 38.29% of the null test statistics were less than $\lambda$, based on the probability that a standard normal observation is between $-1/2$ and $1/2$. The selection of $\lambda$ is a topic of active research (Storey 2002b).

**Computing posterior probabilities**

This approach can also be used to compute the posterior probability that each null hypothesis is true. Eq. (5) has $p$ as a function of the threshold that demarcates the optimal rejection region: $p = P(\tilde{\tau}) = \Pr(H = 0 \,|\, t = \tilde{\tau})$. Thus, as mentioned above, $P(T)$ is the probability that a null hypothesis of observed test statistic $T$ is true. The estimator $\hat{\tau} = \hat{\tau}(p)$, considered above, estimates $\tilde{\tau} = \tilde{\tau}(p) = P^{-1}(p)$, the inverse of $P(T)$. Thus, $P(T)$ can be estimated by $\hat{P}(T)$, an approximate inverse of $\hat{\tau}(p)$. The estimator $\hat{P}(T)$ is defined here as the value of $p$ at which $\hat{\tau}(p)$ is closest to $T$ without exceeding $T$:

$$\hat{P}(T) \equiv \min \{p : T \geq \hat{\tau}(p), p \in \{\partial, 2\partial, ..., 1-\partial\}\}, \tag{18}$$

taking advantage of the monotonicity of $\hat{\tau}(p)$, where $\partial$ is a sufficiently small, positive constant. The computation of $\hat{P}(T)$ for each null hypothesis is similar to Storey's (2002a) computation of a q-value for each null hypothesis in that both computations consider multiple rejection regions. Each approach has a notable advantage over the other. The main advantage of computing the q-values for all null hypotheses is time efficiency, requiring only $m$ optimizations, whereas the computation of $\hat{P}(T)$ for all null hypotheses requires at least $m^2$ optimizations. On the other hand, $\hat{P}(T_i)$ has the simple interpretation as the probability that the $i$th null hypothesis is true, given the data.

**Comparison to density estimation**

Simulations were used to compare the proposed decision-theoretic method of rejecting hypotheses of low posterior probability to a similar method based on the density estimator of Genovese and Wasserman (2002). Each simulated data set consisted of 20 independent observations of each of 5000 variables for each of two groups, a "treatment" group and a "control" group. For each of the first 1000 variables of the treatment group, all observations were drawn either from N(−2, 1) or from N(2, 1), with probability 1/2 of selecting from either distribution, where N($\mu$, $\sigma^2$) is the normal distribution with mean $\mu$ and standard deviation $\sigma$. The other (4000 + 5000) × 20 observations were drawn from N(0, 1), for a total of 5000 × 2 × 20 observations per data set. (This corresponds to 20 microarrays from each group, with 1000 out of 5000 genes differentially expressed between the two groups.) For each pair of 5000 variables, the null hypothesis is that the two variables of the pair are identically distributed. Thus, the null hypothesis is false for the first 1000 variable pairs and true for the other 4000 variable pairs.

For each simulated data set, two empirical Bayes methods were used to determine which of the 5000 null hypotheses would be rejected: a variant of the above decision-theoretic procedure and a density-estimation method, both of which reject null hypotheses with posterior probabilities less than or equal to some value $p$, which was set to 5%, 10%, 20%, 30%, and 50% to determine the effect of $p$ on the relative performance of the two methods. In the notation of the last section, each $T_i$ was set equal to $-\mathcal{P}_i$, where $\mathcal{P}_i$ is the p-value of the $i$th two-sample, equal variance, two-sided t-test, with $\hat{F}_0(\tau) = F_0(\tau) = |\tau| = -\tau$ for $0 \leq -\tau \leq 1$, since the null distribution for p-values is uniform on [0,1]. For simulation studies, the use of p-values is preferred since it avoids the computational time required by resampling. In the decision-theoretic method, a null hypothesis was rejected if and only if its p-value was less than or equal to the threshold $|\hat{\tau}|$, given by

$$|\hat{\tau}| = -\operatorname*{argmax}_{\tau \in \{-\mathcal{P}_1, -\mathcal{P}_2, \ldots, -\mathcal{P}_m\}} \hat{\delta}(\tau), \tag{19}$$

with $m = 5000$ and with $\hat{\delta}(\tau)$ depending on the posterior probability threshold $p$, as per Eq. (13). The density-estimation method instead rejected the $i$th null hypothesis only if $\hat{P}_i$, its estimated posterior probability of being true, satisfied $\hat{P}_i \le p$, or if $\hat{P}_j \le p$ for some $\mathcal{P}_j \le \mathcal{P}_i$,

$$\hat{P}_i = \hat{\pi}_0 \frac{f_0(\mathcal{P}_i)}{\hat{f}(\mathcal{P}_i)} = \frac{\hat{\pi}_0}{\hat{f}(\mathcal{P}_i)}, \tag{20}$$

following Efron et al. (2001) and Genovese and Wasserman (2002). The estimated density $\hat{f}$ was computed using the observed p-values and the default parameters of the function *density* of R version 1.5.1 (www.r-project.org), except with the standardized median absolute deviation instead of the standardized interquartile range and with $1/4$ instead of $1/5$ as the smoothing parameter exponent; this undersmoothing was suggested by Genovese and Wasserman (2002). Fig. 1 displays the estimated density of p-values for the first simulated data set. For each data set and each value of $p$, the Bayes' error (1) was computed for each of the two methods:

$$\operatorname{BE}(\Gamma) = \frac{1-p}{m} \sum_{i=1}^{m} R_i (1 - H_i) + \frac{p}{m} \sum_{i=1}^{m} (1 - R_i) H_i; \tag{21}$$

here, $R_i = 1$ if the $i$th null hypothesis is rejected, $R_i = 0$ if it is not rejected, $H_i = 1$ if $i \le 1000$, and $H_i = 0$ if $i > 1000$.

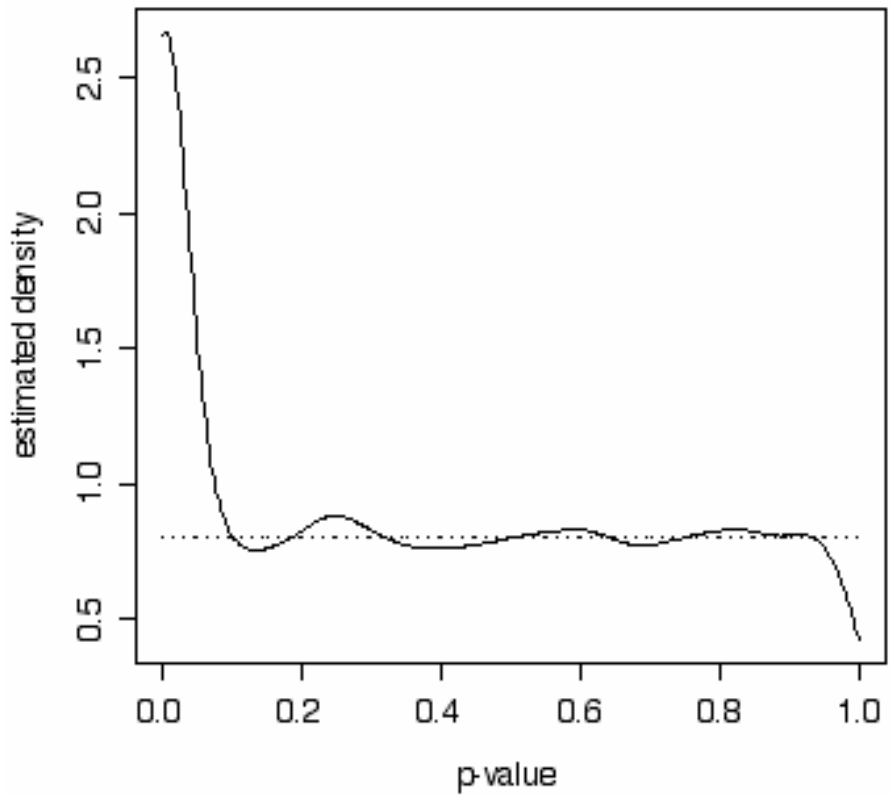

**Estimated probability density of 5000 observed p-values for the first simulated data set.** The dashed line gives the expected contribution of the null distribution to the density (uniform distribution with 80% of the probability). The probability concentrated in the region of lower p-values corresponds to the contribution of the alternate distribution.

**Figure 1**

A total of 20 independent data sets were generated, thereby yielding, at each value of $p$, 20 values of BE for the decision-theoretic method and 20 values of BE for the density-estimation method. The mean BE over the 20 data sets is consistently lower for the former method (Fig. 2), as expected from the unreliably of density estimation (Genovese and Wasserman 2002). For each data set, the ratio of the density-estimation BE to the decision-theoretic BE was computed; these ratios indicate that the two methods have similar performance for $p = 0.2$, but that the decision-theoretic is clearly superior at other values of $p$.

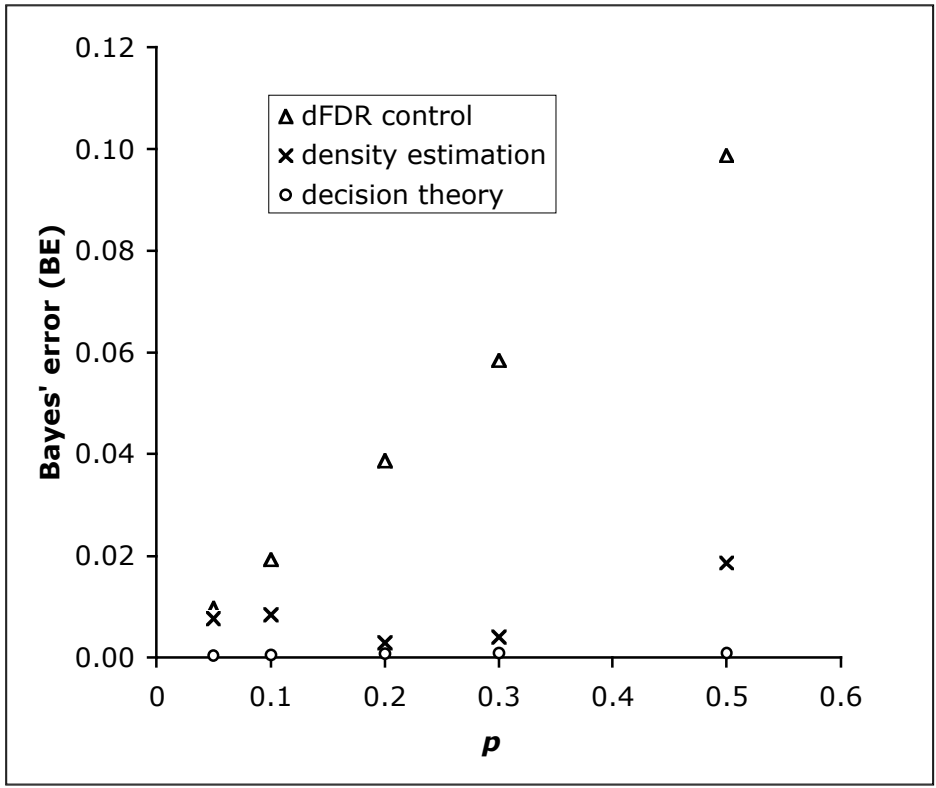

**Bayes' error (21) mean over 20 simulated data sets for three methods of determining which null hypotheses to reject.** The dFDR control method rejects the *i*th null hypothesis if and only if $\mathcal{P}_i \leq {}^\Delta\mathcal{P}$, where ${}^\Delta\mathcal{P}$ is the highest value such that $\hat{\Delta}(-{}^\Delta\mathcal{P}) \leq p$. Its poor performance reflects the fact that it was not designed to minimize the BE. The other two methods, described in the text, are designed to reject null hypotheses with posterior probabilities less than or equal to *p*.

**Figure 2**

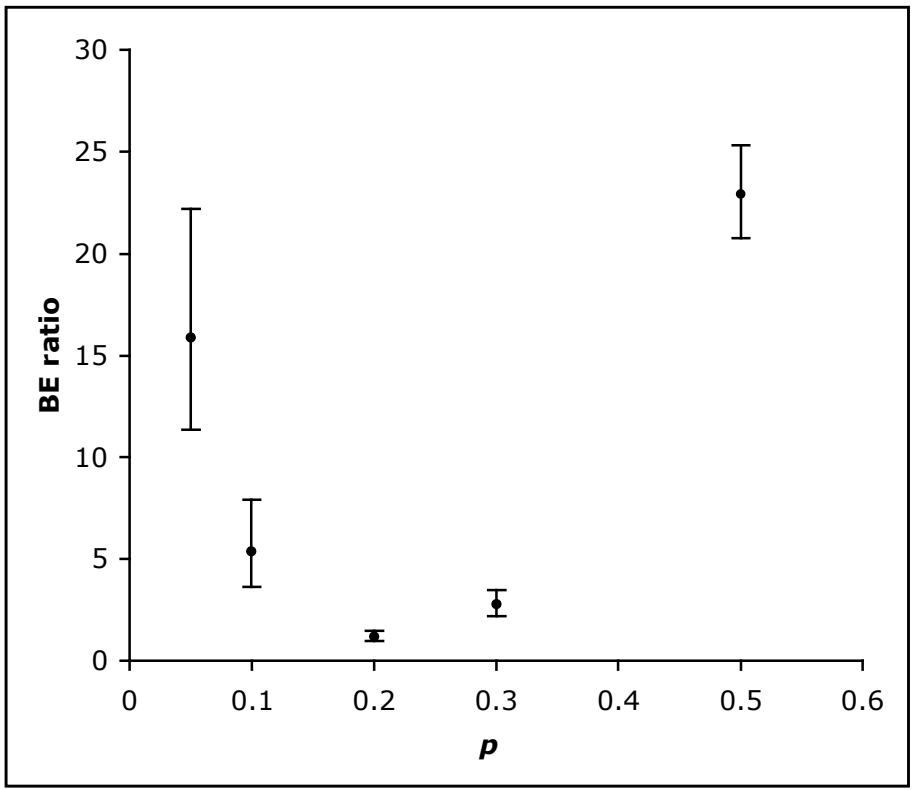

**Geometric means of the ratios of the BE of the density-estimation method to the BE of the decision-theoretic method, over the 20 simulated data sets.** The error bars give approximate 68% confidence intervals. They correspond to the symmetric error bars found using the standard error of the sample mean of the log-transformed ratios.

**Figure 3**

**Application to microarray data**

The same decision-theoretic and density-estimation methods applied to each of the simulated data sets were also applied to the problem of determining which genes are differentially expressed between 38 patients with B-cell acute lymphoblastic leukemia (ALL) and 25 patients with acute myeloid leukemia (AML), based on the expression levels of $m = 7129$ genes; this microarray data set of Golub et al. (1999) is publicly available. Before computing the p-values $(\mathcal{P}_i)_{i=1}^m$, the observed expression values, $\left((x_{i,j})_{j=1}^{38+25}\right)_{i=1}^m$, called "average differences," were transformed as per Bickel (2002, 2003): $x'_{i,j} = \text{sign}(x_{i,j}) \ln(1 + |x_{i,j}/\text{median}_{i'}(x_{i',j})_{i'=1}^m|)$. A gene is considered *differentially expressed* if its corresponding null hypothesis is rejected as improbable.

The normality and equal variance assumptions are very reasonable for the transformed data since almost identical results were obtained without those assumptions, by the unequal-variance, permutation-testing procedure (Table 1). Fig. 4 displays the p-value density that was computed in the density-estimation method and Fig. 5 has the resulting posterior probabilities of differential expression, $1 - \hat{P}_i$. In light of the findings of the simulation study of the previous section and the theoretical study of Genovese and Wasserman (2002), the decision-theoretic analysis of this cancer data set is probably much more reliable than the density-estimation analysis, but results of the latter are shown for comparison in Table 2. It is concluded that decision theory more reliably determines which genes have a high probability of differential expression than does the estimation of probability densities.

| $p=0.05$ | Statistic threshold | Number of discoveries | dFDR |
|---|---|---|---|
| **Proposed method** (equal variance, normal) | 3.12 | 874 | 1.28 % |
| **Permutation method** | 3.14 | 910 | 1.25 % |

**Comparison of two decision-theoretic methods of detecting differential gene expression between ALL and AML for $p = 0.05$.** The proposed method, described in the section on simulations, actually used a p-value threshold, but the corresponding threshold of the absolute value of the t-statistic is given here; a gene with an absolute value of the t-statistic greater than or equal to the threshold was considered differentially expressed and called a *discovery*. The dFDR was estimated by $\hat{\Delta}(\tau)$ of equation (16) and is consistent with the dFDR inequality (10). The values reported for the permutation method are from Bickel (2003). Since lowest value of $\hat{P}_i$ is 0.104 for this data set, the density-estimation method fails to make any discoveries of differential expression for $p<0.104$, and thus is not represented on this table.

**Table 1**

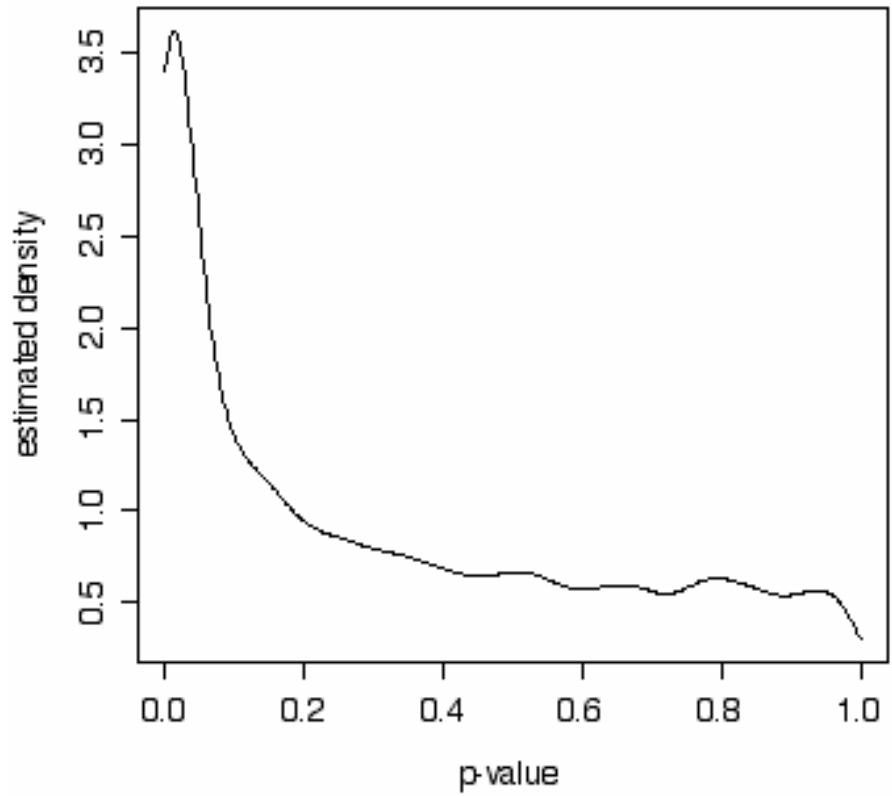

**Estimated probability density for the p-values of the t-tests applied to the leukemia data set.**

**Figure 4**

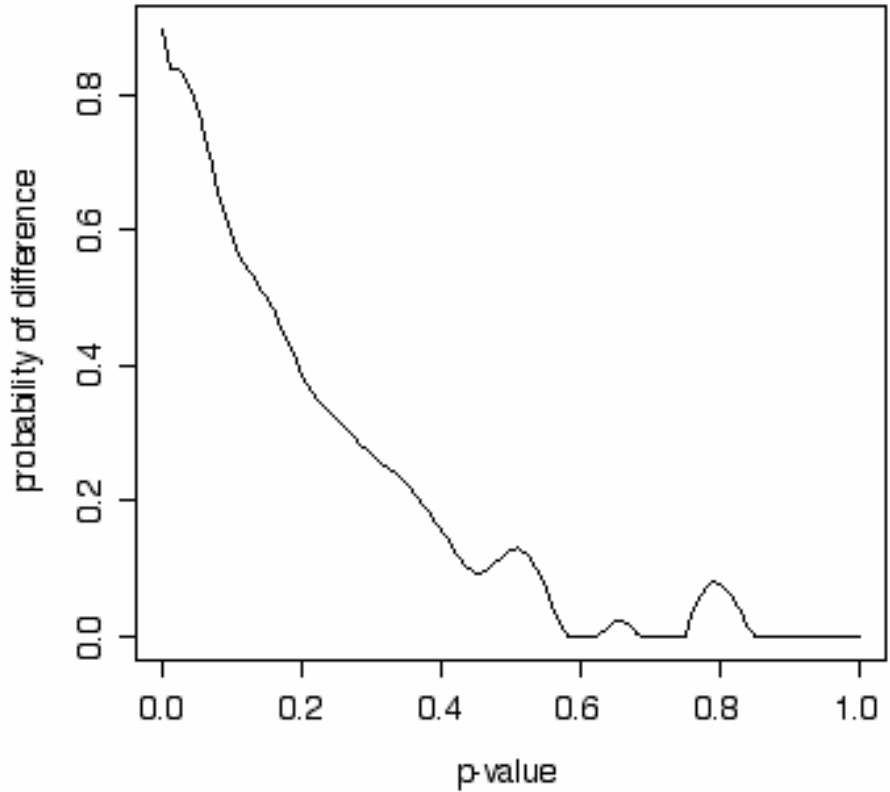

**Posterior probability that a gene of the given p-value is differentially expressed between ALL and AML, computed by the density-estimation method (20).**

**Figure 5**

| $p$=0.20 | Statistic threshold | Number of discoveries | dFDR |
|---|---|---|---|
| **Proposed method** (decision theory) | 2.26 | 1702 | 6.62 % |
| **Density–estimation method** | 2.08 | 1954 | 8.83 % |

**Comparison of the decision-theoretic method to the density-estimation method of detecting differential gene expression between ALL and AML for $p = 0.20$.** The similar values are reminiscent of the similar performance of the two methods in the simulations at $p$=0.20 (Fig. 3).

**Table 2**

# References

Benjamini, Y. and Hochberg, Y. 1995. Controlling the false discovery rate: A practical and powerful approach to multiple testing, *Journal of the Royal Statistical Society* **B 57**, 289-300

Bickel, D. R. 2002. Microarray gene expression analysis: Data transformation and multiple comparison bootstrapping, *Computing Science and Statistics* **34**, 383-400, Interface Foundation of North America (*Proceedings of the 34th Symposium on the Interface*, Montréal, Québec, Canada, April 17-20, 2002); available from www.mathpreprints.com

Bickel, D. R. 2003. Selecting an optimal rejection region for multiple testing: A decision-theoretic alternative to FDR control, with an application to microarrays, submitted; *arXiv.org* e-print math.PR/0212028; available from http://arxiv.org/math.PR/0212028

Bickel, D. R. 2004. Degrees of differential gene expression: Detecting biologically significant expression differences and estimating their magnitudes, *Bioinformatics* (to appear)

Dudoit, S., Yang, Y. H., Speed, T. P., and Callow, M. J. 2002. Statistical methods for identifying differentially expressed genes in replicated cDNA microarray experiments, *Statistica Sinica* **12**, 111-139

Efron, B., Tibshirani, R., Storey, J. D., and Tusher, V. 2001. Empirical Bayes analysis of a microarray experiment, *Journal of the American Statistical Association* **96**, 1151-1160

Genovese, C. and Wasserman, L. 2001. Operating characteristics and extensions of the FDR procedure, Technical report, Department of Statistics, Carnegie Mellon University

Genovese, C. and Wasserman, L. 2002. Bayesian and frequentist multiple testing, unpublished draft of 4/12/02

Golub, T. R., Slonim, D. K., Tamayo, P., Huard, C., Gaasenbeek, M., Mesirov, J. P., Coller, H., Loh, M. L., Downing, J. R., Caligiuri, M. A., Bloomfield, C. D., and Lander, E. S. 1999. Molecular classification of cancer: Class discovery and class prediction by gene expression modeling, *Science* **286**, 531-537

Müller, P., Parmigiani, G., Robert, C., and Rousseau, J. 2004. Optimal sample size for multiple testing: the case of gene expression microarrays (February 2, 2004). Johns Hopkins University, Dept. of Biostatistics Working Papers. Working Paper 31. http://www.bepress.com/jhubiostat/paper31

Storey, J. D. 2002a. A direct approach to false discovery rates, *Journal of the Royal Statistical Society, Series B* **64**, 479-498

Storey, J. D. 2002b. *False Discovery Rates: Theory and Applications to DNA Microarrays*, PhD dissertation, Department of Statistics, Stanford University

Storey, J. D. and Tibshirani, R. 2003. Statistical significance for genome-wide experiments, submitted; available from www.stat.berkeley.edu/~storey/